\documentclass[letter,11pt]{article}
\usepackage{amsmath}
\usepackage{amsthm}
\usepackage{amsfonts}
\usepackage{amssymb}
\usepackage{latexsym}
\usepackage{epsfig,graphicx,psfrag}
\usepackage{enumerate}

\newtheorem{theorem}{Theorem}

\newtheorem{corollary}[theorem]{Corollary}

\newtheorem{lemma}[theorem]{Lemma}
\newtheorem{definition}[theorem]{Definition}
\newtheorem{proposition}[theorem]{Proposition}

\numberwithin{equation}{section}
\newenvironment{Proof}{\noindent\bf{Proof.}\rm}{\hfill$\blacksquare$\bigskip}

\DeclareMathOperator{\diag}{diag}

\newcommand\abs[1]{\left|{#1}\right|}
\numberwithin{theorem}{section}
\newcommand{\kron}{Kronecker}
\newcommand{\sgp}{SGP_{k,t}}

\newcommand{\R}{{\mathbb{R}}}

\newcommand{\C}{{\cal{C}}}
\newcommand{\cC}{{\cal{C}}}
\newcommand{\B}{{\cal{B}}}
\newcommand{\cX}{{\cal{X}}}
\newcommand{\cY}{{\cal{Y}}}
\newcommand{\cB}{{\cal{B}}}

\newcommand{\eps}{{\varepsilon}}

\newcommand{\x}{{\bf x}}
\newcommand{\uu}{{\bf u}}
\newcommand{\vv}{{\bf v}}
\newcommand{\w}{{\bf w}}
\newcommand{\g}{{\bf g}}
\newcommand{\h}{{\bf h}}
\newcommand{\si}{{\bf c}}

\newcommand{\tauu}  {\boldsymbol \tau}
\newcommand{\bee}{{\bf b}}
\newcommand{\one}{{\bf 1}}

\newcommand{\y}{{\bf y}}
\newcommand{\ay}{{\bf a}}

\newcommand{\bgp}{BGP_{k,t}}
\newcommand{\bgpp}[1]{BGP_{k,t,#1}}
\newcommand{\gp}{GP_{k,t}}

\makeatletter
\newcommand*{\rom}[1]{\expandafter\@slowromancap\romannumeral #1@}
\makeatother

\title{Edge distribution in generalized graph products}
\author{Michael Langberg
        \thanks{Department of Mathematics and Computer Science, The Open University of Israel, 108 Ravutski St., Raanana 43107, Israel, email:
            mikel@openu.ac.il. Work suppoted in part by ISF grant 480/08.}\\ \and Dan Vilenchik
        \thanks{Dept. of Computer Science and Applied Math., The Weizmann Institute of Science, Rehovot 76100, Israel,
email dan.vilenchik@weizmann.ac.il. Work done in part while at The Open University of Israel.}}
\date{}

\begin{document}

\maketitle

\begin{abstract} Given a graph $G$ and a natural number $k$, the $k^{th}$ graph product of $G=(V,E)$ is the graph with vertex set $V^k$. For every two vertices $\x=(x_1,\ldots,x_k)$ and $\y=(y_1,\ldots,y_k)$ in $V^k$, an edge is placed according to a predefined rule. Graph products are a basic combinatorial object, widely studied and used in different areas such as hardness of approximation, information theory, etc. We study graph products with the following ``$t$-threshold" rule: connect every two vertices $\x,\y \in V^k$ if there are at least $t$ indices $i \in [k]$ s.t. $(x_i,y_i)\in E$. This framework generalizes the well-known graph {\em tensor-product} (obtained for $t=k$) and the graph {\em or-product} (obtained for $t=1$). The property that interests us is the edge distribution in such graphs. We show that if $G$ has a spectral gap, then the number of edges connecting ``large-enough" sets in $G^k$  is ``well-behaved", namely, it is close to the expected value, had the sets been random.
We extend our results to {\em bi-partite graph} products as well.
For a bi-partite graph $G=(X,Y,E)$, the $k^{th}$ bi-partite graph product of $G$ is the bi-partite graph with vertex sets $X^k$ and $Y^k$ and edges between $\x \in X^k$ and $\y \in Y^k$ according to a predefined rule. 
A byproduct of our proof technique is a new explicit construction of a family of co-spectral graphs.
\end{abstract}

\newpage

\section{Introduction}
Given a graph $G$ and a natural number $k$, the $k^{th}$ graph product of $G=(V,E)$ is the graph with vertex set $V^k$. For every two vertices $\x=(x_1,\ldots,x_k)$ and $\y=(y_1,\ldots,y_k)$ in $V^k$, an edge is placed according to a predefined rule. Graph products are a basic combinatorial object, widely studied and used in different areas such as hardness of approximation, information theory, etc. The graph product operation with parameter $k$ is sometimes referred to as the $k^{th}$ power of $G$.

There are several well-known graph products that have been studied in the literature. The tensor product, introduced by Alfred North Whitehead and Bertrand Russell in their 1912 Principia Mathematica, is defined using the rule ``connect every two vertices $\x,\y \in V^k$ if all indices $i$ satisfy $(x_i,y_i) \in E(G)$". This graph product appears in the literature also as the weak product, or the conjunction product. It is equivalent to the~\kron~product of graph adjacency matrices~\cite{Weichsel62}. A famous standing conjecture related to tensor products is Hedetniemi's conjecture from 1966 \cite{Hedetniemi66}, which states that the chromatic number of the tensor product of two graphs $G_1,G_2$ is at most the minimum between the chromatic numbers of $G_1$ and $G_2$.

The graph or-product (also called the co-normal/disjunctive graph product) is defined using the rule ``connect $\x$ and $\y$ if there exists at least one index $i$ s.t. $(x_i,y_i)\in E$". A slightly different definition of the or-product was used by Garey and Johnson in their famous book \cite{GareyJohnson79} to prove the first hardness of approximation results for the graph coloring problem and the largest independent set problem (actually the book cover itself features a drawing of a graph product). In general, graph products are used as a tool in establishing hardness of approximation results, and specifically in amplifying existing hardness results.

Other types of graph products were studied as well, for example the Xor graph product in which an edge $(\x,\y)$ is present iff
an odd number of coordinates $i$ satisfy $(x_i,y_i) \in E$. This product was studied in \cite{Thomason97,AlonLube07}.
One can generalize the definitions above by taking the product of $k$ different graphs $G_1,G_2,\ldots,G_k$.


\subsection{Our Contribution and Techniques}
We study two natural generalizations of the ``or'' and ``tensor'' products. Observe that the only difference in the definition of the two products is in the edge-threshold parameter $t$: at least one index $i$ satisfying $(\x_i,\y_i)\in E$ for the or-product, and all $k$ indices satisfying $(\x_i,\y_i) \in E$ for the tensor product. In this work we study graph products with an arbitrary threshold $t$. Using this notation, the tensor product is obtained by taking $t=k$ and the or-product by $t=1$. Our notion of graph product is similar in spirit to the special graph products known as the non-complete extended $P$-sum (NEPS) graph products, introduced for the first time by Cvetkovi$\rm \acute{c}$ in the early 1970's \cite{Cvetkovi71}, and elaborated in \cite{Cvetkovi95}. One generalization of NEPS graph products
given in \cite{Sokarovski77} actually captures our $t$-threshold graph product rule as a special case. More details are given in the technical sections.

The second generalization that we suggest is a {\em bi-partite} graph product with threshold $t$. Specifically, let $G=(X,Y,E)$ be a bi-partite graph, and define the $k^{th}$ bi-partite power of $G$ to be the graph with vertex sets $X^k \cup Y^k$ and edges according to some predefined rule.


In this work we consider only regular graphs $G$, and study the edge distribution in the $k^{th}$ power of $G$. A well-known edge discrepancy result  for regular graphs is the \emph{Expander Mixing Lemma} \cite{EML88} (EML for short). The EML states that for a $d$-regular $n$-vertex graph $G$, with second largest eigenvalue (in absolute value) $\lambda$, and for any two sets of vertices $S,T$, the number of edges they span in $G$, $e(S,T)=\{(u,v) \in E \mid u \in S, v \in T\}$ (if $S$ and $T$ are not disjoint, then edges in the intersection are counted twice), satisfies
\begin{equation}\label{eq:EML}
\left|e(S,T)-\frac{d|S||T|}n\right| \leq \lambda \sqrt{|S||T|}.
\end{equation}
The quantity $d|S||T|/n$ can be thought of as the expected number of edges spanned by random sets $S$ and $T$. The intuitive way to read the EML is that if $\lambda < d$, then the number of edges between ``large enough" sets is well-behaved, namely it is close to the expected value. The EML as stated gives a trivial bound for $d$-regular bi-partite graphs $G$ (since $\lambda(G) = |-d| = d$ as well). One can rather easily extend and restate the EML for $d$-regular $n$-vertex bi-partite graphs (in which case $\lambda$ in Eq.~(\ref{eq:EML}) refers to the third largest eigenvalue in absolute value, and $d|S||T|$ is divided by $n/2$, which is the number of vertices in each partition, due to regularity):
\begin{equation}\label{eq:EMLbip}
\left|e(S,T)-\frac{d|S||T|}{n/2}\right| \leq \lambda \sqrt{|S||T|}.
\end{equation}

Our study focuses on applying the Expander Mixing Lemma to our generalized graph products. The main challenge, after showing that $G^k$ is regular and computing its degree, is to obtain an upper bound on $\lambda$.
The spectrum of the tensor product, i.e. the case $t=k$, is well-understood, and has a simple clean characterization~\cite{Laub04}. For $t < k$ one has to work harder to compute $\lambda$.
The more challenging case is the bi-partite graph product. While previous results provide access to the spectrum of the non-bipartite graph product, no such result exists for the bi-partite product. The main technical achievement of this paper is giving an exact characterization of the spectrum of $G^k$ in the bi-partite setting. This is obtained by embedding $G^k$ in a certain larger graph, for which the spectrum can be analyzed, and then infer the spectrum of $G^k$ from that graph. Let us remark that for both types of graph products, optimality results can be obtained from the ``converse to the EML" result obtained in \cite{BiluLinial06}.

Another way of viewing our results is using the terminology of \textbf{$(d,\alpha)$-jumbled graphs}, coined in \cite{Thomason87}, and used widely in the study of pseudo-random graphs (see the excellent survey in \cite{KrivSudakok06} for more details). A $d$-regular graph $G$ is called $(d,\alpha)$-jumbled if any two sets of vertices $S,T$ satisfy Eq.~(\ref{eq:EML}), with $\alpha$ replacing $\lambda$. We compute the value $\alpha$, s.t. $G^k$ is $(d',\alpha)$-jumbled ($d'$ is the degree of $G^k$). We do that for both types of graph products, and for every parameter $t$.

Finally let us mention a byproduct of our proof technique, which may be of self interest. Recall that the spectrum of a graph $G$ is the multiset of eigenvalues of its adjacency matrix. Two graphs
are called \textbf{cospectral} if they have the same spectrum. Clearly isomorphic graphs are cospectral, but the interesting question is to identify pairs of graphs which are not isomorphic, yet cospectral.
There are numerous constructions of such pairs. Seidel switching, introduced in \cite{LintSeidel66}, is a quite
general method to construct such pairs, and was used in \cite{GodsilMcKay82} for example. In this paper we give a new construction of a family of co-spectral graphs. Our construction is natural and simple to describe, and uses a different approach, which is closer in flavor to the construction in~\cite{GodsilMcKay76}. Instead of local switching, our technique can be viewed as a family of blow-ups of a $d$-regular bi-partite graph $G$ that produce a family of co-spectral graphs.

To prove the co-spectrality result, we prove an auxiliary technical lemma which may be of self interest (Lemma \ref{lem:NumOfTemplatePathsInG}). Given a bi-partite graph $G=(X\cup Y,E)$ and a vector $\uu=(u_1,\ldots,u_\ell) \in \{0,1\}^\ell$, we define a closed $(\uu,X)$-walk in $G$ as an alternating walk $(x_{i_1},y_{i_2},\ldots,x_{i_\ell}=x_{i_1})$ of (even) length $\ell$ starting at some vertex $x_{i_1} \in X$ (and ending at the same vertex), in which the $i^{th}$ vertex along the walk belongs to the neighbors of vertex $i-1$ if $u_{i-1}=1$ and to its non-neighbors if $u_{i-1}=0$. For $\uu=\overrightarrow{\textbf{1}}$, we get the standard notion of a walk. We prove that for every $d$-regular bi-partite graph $G$, for every $\ell$ and for every vector $\uu \in \{0,1\}^\ell$, the number of closed $(\uu,X)$-walks equals the number of closed $(\uu,Y)$-walks.

We proceed with a formal statement of our results in Section \ref{sec:Results}. All the technical details and proofs follow in subsequent sections.



\section{Main Results}\label{sec:Results}
We start with the formal definition of the graph product. For sake of precision, we replace the notation ``$\ G^k\ $'', which we used in the Introduction, with a more detailed and explicit notation.

\begin{definition} \label{defn:GraphProduct}
For a simple graph $G=(V,E)$, define the \textbf{graph product} $\gp(G)$ to be the graph with vertex set $V^k$, and an edge $(x_1,\ldots,x_k)\sim(y_1,\ldots,y_k)$ if there are at least $t$ pairs $(x_i,y_i)$ that share an edge in $G$.
\end{definition}

\begin{definition} \label{defn:BipGraphProduct}
If $G=(X,Y,E)$ is a simple bipartite graph, define the \textbf{bi-partite graph product} $\bgp(G)$ to be the graph with vertex set $(X^k,Y^k)$, and an edge between $(x_1,\ldots,x_k) \in X^k$ and $(y_1,\ldots,y_k) \in Y^k$ if there are at least $t$ pairs $(x_i,y_i)$ that share an edge in $G$.
\end{definition}

Let us first establish the easy fact that both graph products are regular graphs, and state their degrees.

\begin{proposition}\label{prop:deg}  Let $G$ be an $n$-vertex $d$-regular graph, and $G'$ be an $n$-vertex bi-partite $d$-regular graph.
Then $\gp(G)$ is $d_1$-regular and $\bgp(G')$ is $d_2$-regular with $d_1$ and $d_2$ given by:
\begin{equation}\label{eq:deg}
d_1=\sum_{t'=t}^k \binom {k}{t'}d^{t'}(n-d)^{k-t'}, \qquad d_2=\sum_{t' =t}^k \binom {k}{t'}d^{t'}\left(\frac{n}{2}-d\right)^{k-t'}.
\end{equation}
\end{proposition}
The proof of Proposition \ref{prop:deg} is rather straightforward and is given in Section \ref{sec:degProof}.

Before we state our main results, we introduce some notations: We let $e(S,T)$ stand for the number of edges connecting two sets of vertices $S$ and $T$ (when the underlying graph is clear from context. If the two sets $S$ and $T$ are not disjoint, then edges in the intersection are counted twice). For a graph $G$, we let $A(G)$ be its adjacency matrix. Somewhat abusing notations, when we refer to graph eigenvalues, we mean the eigenvalues of its adjacency matrix. For a graph $G$ with eigenvalues $\lambda_1 \geq \cdots \geq \lambda_n$, let $\lambda(G)=\max\{\lambda_2,|\lambda_n|\}$. If $G$ is bi-partite then  $\lambda(G) = \max\{\lambda_2,|\lambda_{n-1}|\}$ (the reason is that $\lambda_1=|\lambda_n|=d$ if $G$ is bi-partite). For a $D$-regular $N$-vertex graph, and for any two sets of vertices $S,T$, we denote by $\mu_{S,T}$ the expected number of edges that connect $S$ and $T$, had the sets been chosen uniformly at random. The following equations give this quantity for the general and bi-partite case:
\begin{equation}\label{eq:MuST}
\mu_{S,T} = \frac{D|S||T|}{N}, \qquad \mu_{S,T} = \frac{D|S||T|}{N/2}=\frac{2D|S||T|}{N}.
\end{equation}

\medskip
Our first theorem provides a tight estimate of the second largest eigenvalue of $\gp(G)$. In the statement of our results we use the following function $\alpha=\alpha(k,t)$:
\begin{equation}\label{eq:Alpha}
\alpha = \sum_{\ell=0}^{k-t}\frac{\binom{k-t}{\ell}}{\binom{t+\ell}{\ell}}.
\end{equation}
Note that $\alpha$ is decreasing with $t$, from $\alpha = O(2^k/k)$ for $t=1$ down to $\alpha =1 $ for $t=k$.

\begin{theorem}\label{thm:GP} Let $\gp(G)$ be the graph product of an $n$-vertex $d$-regular graph $G=(V,E)$ with $\lambda=\lambda(G)$.
Let $\Lambda=\lambda(\gp(G))$. If $d \le (n-1)/2$  then
$$\frac{\lambda}{d}\cdot \frac{1}{\alpha}\cdot \frac{t}{k} \cdot d_1\leq \Lambda \leq  \frac{\lambda}{d}\cdot \frac{1}{\alpha} \cdot d_1.$$
\end{theorem}
Theorem \ref{thm:GP} is derived from \cite[Thm 3]{Sokarovski77}, where a general formula for the eigenvalues of a large class of graph products was established. Obtaining a tight and succinct estimate on $\Lambda$, as we did, requires additional work. The following result about the  edge-distribution in the non-bipartite graph product is an immediate corollary of Theorem \ref{thm:GP} and the EML (Eq.~(\ref{eq:EML})):
\begin{corollary}\label{cor:GP} Under the conditions of Theorem \ref{thm:GP}, for any two sets of vertices $S,T$ in $\gp(G)$
\begin{equation}\label{eq:thm1}
\left|e(S,T)-\mu_{S,T}\right| \leq  \frac{1}{\alpha}\cdot \frac{\lambda}{d}\cdot d_1\sqrt{|S||T|}.
\end{equation}
\end{corollary}

The next theorem is the analogue of Theorem \ref{thm:GP} for the bi-partite setting. While the adjacency matrix of $\gp(G)$ has an explicit closed expression (based on matrix tensor products), this is not the case for $\bgp(G)$. Therefore the task of estimating $\Lambda$ is more complicated.

Recall that for a bi-partite graph $G$, $\lambda(G)$ is the third largest eigenvalue in absolute value.
\begin{theorem}\label{thm:BiPartGP} Let $\bgp(G)$ be the bi-partite graph product of an $n$-vertex bi-partite $d$-regular graph $G=(X,Y,E)$ with $\lambda=\lambda(G)$. Let $\Lambda$ be the third largest eigenvalue in absolute value of $\bgp(G)$. If $d \le n/4$ then

$$\frac{\lambda}{d}\cdot \frac{1}{\alpha}\cdot \frac{t}{k}  \cdot d_2 \le\Lambda \le \frac{\lambda}{d} \cdot d_2.$$
\end{theorem}

Consequently, together with the EML for bi-partite graphs given in Eq.~(\ref{eq:EMLbip}), we get the following edge-distribution result for $\bgp(G)$:
\begin{corollary}\label{cor:BiPartGP} Under the conditions of Theorem \ref{thm:BiPartGP}, for any two sets of vertices $S \subseteq X^k$ and $T \subseteq Y^k$,
\begin{equation}\label{eq:thm1bip}
\left|e(S,T)-\mu_{S,T}\right| \leq  \frac \lambda {d}\cdot d_2\sqrt{|S||T|}.
\end{equation}
\end{corollary}

\noindent A few remarks concerning Theorems \ref{thm:GP} and \ref{thm:BiPartGP} are in place.

\begin{enumerate}
\item Our estimate on $\Lambda$ in Theorem \ref{thm:GP} is tight, up to a factor of $t/k$ (which for most $t$ values is negligible compared to the much bigger $\alpha$).
If $t=k$, namely $\gp(G)$ is the tensor product, then our lower and upper bound on $\Lambda$ coincide, and $\alpha=1$. This is of course inline with the well known result for the tensor product stating that $\lambda(\gp(G)) =  \lambda d^{k-1}$ (obtained from our bound by setting $d_1=d^k$). For the bi-partite case, Theorem \ref{thm:BiPartGP}, we are missing a $1/\alpha$ factor in the upper bound on $\Lambda$, which is an artifact of our proof technique. Nevertheless, for the bi-partite tensor product ($t=k$), the lower and upper bound on $\Lambda$ coincide.

\item The upper bound on the discrepancy in Corollaries \ref{cor:GP} and \ref{cor:BiPartGP} is tight (up to $\log d$-factors, and factors that depend on $k$). This follows from the converse to the EML result proven in \cite{{BiluLinial06}}, together with our lower bound on $\Lambda$ in Theorems \ref{thm:GP} and \ref{thm:BiPartGP}. The converse to the EML reads as follow: Let $H$ be an $N$-vertex $D$-regular graph. If for any $S,T$ with $S\cap T=\emptyset$
$\left|e(S,T)- \mu_{S,T}\right| \leq  \alpha \sqrt{|S||T|}$, then all but the largest eigenvalue of $H$ are bounded in absolute value by $O(\alpha(1+\log(D/\alpha)))$. A similar result is derived in \cite{BiluLinial06} for a bi-partite regular graph, in which case the same upper bound is obtained on all but the two largest eigenvalues.
\item We can restate Theorems \ref{thm:GP} and \ref{thm:BiPartGP} in the jumbled-graph terminology~\cite{Thomason87}, namely $\gp(G)$ is $(d_1,\frac{1}{\alpha}\cdot\frac{\lambda}{d}\cdot d_1)$-jumbled, and $\bgp(G)$ is $(d_2,\frac{\lambda}{d}\cdot d_2)$-jumbled.

\item The assumptions on $d$ in the statement of both theorems are made to enable a succinct description for the estimates on $\Lambda$. Eq.~(\ref{eq:ExactFormula})~and~(\ref{eq:ExactFormulaBip}) spell an exact expression for $\Lambda$.
\end{enumerate}

\medskip

Finally, let us demonstrate an \textbf{application} of Theorem \ref{thm:GP}. Let $G$ be a $d$-regular expander graph with $\lambda \leq 2\sqrt{d}$. Consider two sets of vertices $S,T$ in $\gp(G)$, and for simplicity assume they have the same size $|S|=|T|=\xi n^k$ for some $\xi \in [0,1]$. For our choice of parameters, the right hand side of Eq.~(\ref{eq:thm1}) is at most
$$\frac{1}{\alpha}\cdot\frac{\lambda}{d} \cdot d_1\sqrt{|S||T|} \le \frac{2}{\sqrt{d}}\cdot d_1|S| = \frac{2}{\sqrt{d}}\cdot \frac{d_1|S||T|}{n^k}\cdot \frac{n^k}{|T|}
=\frac{2}{\xi\sqrt{d}}\cdot\mu_{S,T}.$$
If, say, $\xi \geq \log d/\sqrt{d}$, then the latter together with Eq.~(\ref{eq:thm1}) imply that
$$e(S,T)=(1 \pm \eps)\frac{d_1|S||T|}{n^k},$$

where $\eps \to 0$ as $d \to \infty$. In words, all ``sufficiently large" sets of vertices have the ``right" number of edges between them.

\subsection{A family of co-spectral graphs}\label{sec:Cospectral}
In Definition \ref{defn:BipGraphProduct} we specified the bi-partite product $\bgp(G)$ of a graph $G=(X,Y,E)$ as a graph over the vertex set $X^k \cup Y^k$.
Namely, each vertex in $\bgp(G)$ is either a $k$-tuple consisting of elements from $X$, or a $k$-tuple consisting of elements from $Y$.
In our proofs to follow, we study additional forms of bipartite graph products for $G$ in which vertices in the product graph are $k$-tuples which may consist of a mixture of elements from both $X$ and $Y$.

Formally, let $\cX$ and $\cY$ be symbols representing the sets $X$ and $Y$ respectively.
We define the concept of a \textbf{template} $\tauu=(\tau_1,\ldots,\tau_k) \in \{\cX,\cY\}^k$, which is a string of length $k$  consisting of $\cX$ and $\cY$ symbols.

A $k$-tuple of vertices $\ay = (a_1,\ldots,a_k) \in (X \cup Y)^k$ is said to have template $\tauu$ iff $a_i \in X$ whenever $\tau_i=\cX$ (and equivalently, $a_i \in Y$ whenever $\tau_i=\cY$).
For a template $\tauu$, we define its complement template $\tauu^c$ by switching $\cal{X}$ to $\cal{Y}$ and vice versa. For example, all vertices in $\bgp(G)$ have either the template $\tauu=\cX^k$ or its complement $\tauu^c=\cY^k$.

For a given template $\tauu$, denote the $k$-tuples in $(X \cup Y)^k$ with template $\tauu$ by $V_{\tauu}$. Now, given a bi-partite graph $G=(X\cup Y,E)$, and a template $\tauu$, we denote by $\bgpp{\tauu}$ the bi-partite graph product with vertex set $V_{\tauu} \cup V_{\tauu^c}$, in which two vertices $\ay$ and
$\bee$ are connected iff at least $t$ pairs $(a_i,b_i)$ are edges in $G$. The graphs $\bgpp{\tauu}$ are used in the proof of Theorem \ref{thm:BiPartGP}. On the way, we prove that

\begin{theorem}\label{thm:cospectral} Let $G=(X \cup Y,E)$ be a $d$-regular graph. Then for every $k \ge 1$ and any two templates $\tauu,\tauu'$, the graphs $\bgpp{\tauu}$ and $\bgpp{\tauu'}$ are co-spectral.
\end{theorem}

Note that any two isomorphic graphs are co-spectral (since their adjacency matrices are permutations of each other, and such matrices have the same spectrum). It is not hard to see that $\bgpp{\tauu}$ and $\bgpp{\tauu'}$ are isomorphic if the number of $\cX$'s is the same in both templates (the isomorphism permutes the vertices in each tuple to match the other template). However, if the number of $\cX$'s is different, then the two graphs are not necessarily isomorphic. We verified this using a computer program. Specifically, we generated a random 3-regular bi-partite graph $G$ on 12 vertices. Our choice of parameters was $k=3$, $t=1$, $\tauu = \cX \cX \cX$, and $\tauu' = \cX \cY \cX$.
For $G$ with these parameters, we computed the adjacency matrices $A_{\tauu}$ of $\bgpp{\tauu}$ and $A_{\tauu'}$ of $\bgpp{\tauu'}$, and compared the two vectors $\diag(A_{\tauu}^4)$ and $\diag(A_{\tauu'}^4)$. The $i^{th}$ entry of each vector is the number of closed walks of length four that start at $x_i$ (or $y_{\frac{n}{2}-i}$ if $i > n/2$). We sorted the two vectors and found them to be different. This clearly excludes the possibility that $\bgpp{\tauu}$ and $\bgpp{\tauu'}$ are isomorphic.

\noindent The proof of Theorem \ref{thm:cospectral} is self contained and is given in full in Section \ref{sec:ProofThmCospectral}.

\section{Proof of Theorem \ref{thm:GP}}\label{sec:ProofGP}
Let $G=(V,E)$ be a $d$-regular $n$-vertex graph with eigenvalues $\lambda_1 \ge \lambda_2 \ge \cdots \ge \lambda_n$, and corresponding eigenvectors $\uu_1,\ldots,\uu_{n}$. Similarly let $\Lambda_1, \Lambda_2,\cdots, \Lambda_{n^k}$ be the eigenvalues of $\gp(G)$, with corresponding eigenvectors $\w_1,\ldots,\w_{n^k}$. In a slightly different setting and using different terminology, $\gp(G)$ was studied in \cite{Sokarovski77} and the following result regarding its spectrum was obtained (we rephrase that result to match our terminology). In what follows, for two vectors $\uu,\vv \in \R^n$, $\uu \otimes \vv$ stands for the standard tensor product of two vectors.

\begin{theorem}\label{thm:Sokarovski}\cite[Thm 3]{Sokarovski77} Let $G$ be an $n$-vertex $d$-regular graph with eigenvalues $d=\lambda_1 \ge \lambda_2 \ge \ldots \ge \lambda_n$, and corresponding eigenvectors $\uu_1,\uu_2,\ldots,\uu_n$. Define $\lambda^*_1=n-1-d$, and
$\lambda^*_i = -1-\lambda_i$ for all $i \geq 2$. Let $\B$ be the set of all vectors in $\{-1,0,1\}^k$ having at least $t$ 1-entries. Then $\gp(G)$ has a spectrum consisting of eigenvalues
\begin{equation}\label{eq:Sokarovski}
\Lambda_{i_1i_2\cdots i_k}=\sum_{\bee \in \B} \prod_{j=1}^k \left(\frac{1+b_j}{2}(\lambda_{i_j})^{|b_j|}+\frac{1-b_j}{2}(\lambda^*_{i_j})^{|b_j|}\right), \qquad i_1,\ldots,i_k \in \{1,\ldots,n\}.
\end{equation}
The corresponding eigenvector is $\w_{i_1 i_2\cdots i_k}=\uu_{i_1}\otimes \uu_{i_2}\cdots \otimes \uu_{i_k}$.
\end{theorem}

Our proof outline is as follows: we first show that $\Lambda_{11\cdots 1}$ is the largest eigenvalue of $\gp(G)$. We then upper bound $\Lambda =\lambda(\gp(G))$, which is the maximum over all $\Lambda_{i_1i_2\cdots i_k}$'s in which for some $j$, $i_j \ne 1$.

To prove that $\Lambda_{11\cdots 1}$ is the largest eigenvalue of $\gp(G)$ it suffices to show that it equals $d_1$, as it is well-known that the largest eigenvalue of a $D$-regular graph is simply $D$. For $\bee \in \cB$ and $s \in [1,k]$, let $\Psi_{s,\bee}=\prod_{j=s}^{\#\bee} \left(\frac{1+b_j}{2}d^{|b_j|}+\frac{1-b_j}{2}(n-d-1)^{|b_j|}\right)$ (here $\#\bee$ is the length of the vector $\bee$ which, as seen below, will change throughout the proof).
Using Eq.~(\ref{eq:Sokarovski}) and the definition of $\lambda_1=d$ and $\lambda_1^* = n-d-1$, we get the following expression for $\Lambda_{11\cdots 1}$:

\begin{equation}\label{eq:Lambda1111}
\Lambda_{11\cdots 1} = \sum_{\bee \in \B} \Psi_{1,\bee}.
\end{equation}

Next we show that~(\ref{eq:Lambda1111}) equals $d_1$. This calculation was carried in \cite{Sokarovski77}, and we present it here for the sake of completeness. Fix a vertex $\x=(x_1,\ldots,x_k)$, and let us compute the number of vertices $\y=(y_1,\ldots,y_k)$ adjacent to $\x$ with respect to a fixed $\bee \in \B$ according to the following rules: if $b_i=1$ then $(x_i,y_i)$ is an edge in $E$ (in this case there are $d$ choices for $y_i$); if $b_i=0$ then $x_i=y_i$ (one choice for $y_i$); if $b_i=-1$ then $x_i\ne y_i$ and $(x_i,y_i) \notin E$ ($n-d-1$ choices for $y_i$). This is reflected exactly in the product in Eq.~(\ref{eq:Lambda1111}). Since every vector $\bee$ specifies a different configuration of neighbors/non-neighbors of $\x$, and $\B$ contains all possible configurations, we obtain $\Lambda_{11\cdots 1} = d_1$.

Finally we compute $\Lambda$. Using our last insight about $\Lambda_{11\cdots 1}$, the largest eigenvalue of $\gp(G)$, we may conclude that $\Lambda$ is the maximum over all $\Lambda_{i_1i_2\cdots i_k}$ in which at least one $i_j$ satisfies $i_j > 1$. Intuitively, for every $i_j > 1$, we loose a factor proportional to $\lambda_{i_j} / d$ in (\ref{eq:Sokarovski}). Therefore, we expect $\Lambda$ to correspond to  $\Lambda_{i_1i_2\cdots i_k}$ where exactly one $i_j > 1$. Let us start by computing the value of such eigenvalues, and then explain why indeed they give $\Lambda$. By symmetry it doesn't matter which $i_j > 1$, so let's assume $i_1 > 1$.

Fix $i >1$ and look at $\Lambda_{i11\cdots 1}$. Denote by $\cC_{k,r} \subseteq \{1,0,-1\}^k$ the set of all vectors with exactly $r$ one entries and by $\cC^{+}_{k,r}$ the set of vectors with at least $r$ one entries. From Eq.~(\ref{eq:Sokarovski}) we derive
\begin{align*}
\Lambda_{i11\cdots 1} = & \sum_{\bee \in \B,b_1=1}\lambda_i \Psi_{2,\bee}+\sum_{\bee \in \B,b_1=0} \Psi_{2,\bee}+\sum_{\bee \in \B,b_1=-1}(-\lambda_i-1) \Psi_{2,\bee}=\\&
\sum_{\si \in \C^+_{k-1,t-1}}\lambda_i \Psi_{1,\si}+\sum_{\si \in \C^+_{k-1,t}} \Psi_{1,\si}+\sum_{\si \in \C^+_{k-1,t}}(-\lambda_i-1) \Psi_{1,\si}=
\sum_{\si \in \C^+_{k-1,t-1}}\lambda_i\Psi_{1,\si}-\sum_{\si \in \C^+_{k-1,t}}\lambda_i\Psi_{1,\si}.
\end{align*}
Simplifying further we obtain the following expression for $\Lambda_{i11\cdots 1}$:
\begin{equation}\label{eq:ExactFormula}
\Lambda_{i11\cdots 1}= \lambda_i\sum_{\si \in \cC_{k-1,t-1} }\Psi_{1,\si}.
\end{equation}
For $\ell \geq 0$ and $\si \in \cC_{k,r+\ell}$, define the set $\varphi_{\ell}(\si)$ consisting of all elements $\si' \in \cC_{k,r}$ that can be obtained by choosing $\ell$ one entries in $\si$ and switching them to zero.
Observe that (\romannumeral 1 \relax) If $\si' \in \varphi_{\ell}(\si)$ then $\Psi_{1,\si} \ge \Psi_{1,\si'}$ (since we switch a `1' to `0');
(\romannumeral 2 \relax) for $\si' \in \cC_{k,r}$ it holds that $\abs{\varphi_{\ell}^{-1}(\si')}= \binom{k-r}{\ell}$;
and (\romannumeral 3 \relax) for $\si \in \cC_{k,r+\ell}$ it holds that $\abs{\varphi_{\ell}(\si)}=\binom{r+\ell}{\ell}$. We conclude that
\begin{align*}
d_1&=\Lambda_{11\cdots 1}=\sum_{\bee \in \B,b_1 = 1 }d\Psi_{2,\bee} + \sum_{\bee \in \B,b_1 = -1 }(n-d-1)\Psi_{2,\bee}+\sum_{\bee \in \B,b_1 = 0}\Psi_{2,\bee} \geq \sum_{\bee \in \B,b_1 = 1 }d\Psi_{2,\bee}\\
&=d\sum_{\ell=0}^{k-t}\left(\sum_{\si \in \C_{k-1,(t-1)+\ell}}\Psi_{1,\si}\right)
\underbrace{\ge}_{\rm(i)-(iii)}
d\sum_{\ell=0}^{k-t}\left(\frac{\binom{k-t}{\ell}}{\binom{t-1+\ell}{\ell}}\sum_{\si \in \C_{k-1,t-1}}\Psi_{1,\si}\right)=\\&=
d\left(\sum_{\si \in \C_{k-1,t-1}}\Psi_{1,\si}\right)\sum_{\ell=0}^{k-t}\frac{\binom{k-t}{\ell}}{\binom{t-1+\ell}{\ell}} \ge d\left(\sum_{\si \in \C_{k-1,t-1}}\Psi_{1,\si}\right)\sum_{\ell=0}^{k-t}\frac{\binom{k-t}{\ell}}{\binom{t+\ell}{\ell}}\underbrace{=}_{(\ref{eq:Alpha}),(\ref{eq:ExactFormula})}d\alpha \cdot\frac{\Lambda_{i1\cdots 1}}{\lambda_i}.
\end{align*}
Rearranging, we get
\begin{equation}\label{eq:UpBoundOnLamb}
\abs{\Lambda_{i11\cdots 1}} \le \frac{d_1\abs{\lambda_i}}{\alpha d}.
\end{equation}
We now turn to lower bound $\abs{\Lambda_{i11\cdots 1}}$.
As before let $\ell \geq 0$.
Let $\si \in \cC_{k,r+\ell}$.
Let $\phi_{\ell}(\si)$ be the set consisting of all elements $\si' \in \cC_{k,r}$ that can be obtained by choosing $\ell$ one entries in $\si$ and switching them to -1.
By our assumption $d \le n-d-1$ in Theorem \ref{thm:GP}, we have (\romannumeral 4 \relax) for $\si' \in \phi_\ell(\si)$: $\Psi_{1,\si} \le \Psi_{1,\si'}$. Consequently,
\begin{align*}
d_1
&=
\Lambda_{11\cdots 1}
=
\sum_{\bee \in \B}\Psi_{1,\bee}
=
\sum_{\ell=0}^{k-t}\left(\sum_{\si \in \cC_{k,t+\ell}}\Psi_{1,\si}\right)
\underbrace{\le}_{\rm(i)-(iv)}
\sum_{\ell=0}^{k-t}\left(\frac{\binom{k-t}{\ell}}{\binom{t+\ell}{\ell}}\sum_{\si \in \C_{k,t}}\Psi_{1,\si}\right)= \\
&= \alpha \left(\sum_{\si \in \C_{k,t}}\Psi_{1,\si}\right)
= \frac{\alpha dk}{t}\left(\sum_{\si \in \C_{k-1,t-1}}\Psi_{1,\si}\right) = \frac{\alpha d}{\lambda_i}\frac{k}{t} \Lambda_{i11\cdots 1}.
\end{align*}
Rearranging we get,
\begin{equation}\label{eq:LowBoundOnLamb}
\abs{\Lambda_{i11\cdots 1}} \geq  \frac{d_1\abs{\lambda_i}}{\alpha d}\cdot \frac{t}{k}.
\end{equation}
All in all, 
$$
\frac{d_1\abs{\lambda_i}}{\alpha d}\cdot \frac{t}{k} \leq \abs{\Lambda_{i11\cdots 1}} \leq  \frac{d_1\abs{\lambda_i}}{\alpha d}.
$$

Finally, consider the case where more than one $i_j$ is greater than 1. Similar arguments to the ones above imply
for example that $|\Lambda_{i_1i_211\cdots 1}|$, for both $i_1,i_2 > 1$, is smaller than $|\Lambda_{i_111\cdots 1}|$ by a factor of $|\lambda_{i_2}|/d$. This together with Eq.~(\ref{eq:ExactFormula}) imply that $\Lambda = |\Lambda_{i^*11\cdots 1}|$ for $i^*$ s.t. $\lambda_{i^*}=\lambda(G)$ (where $i^*$ is either 2 or $n$).
Theorem \ref{thm:GP} follows from Eq.~(\ref{eq:UpBoundOnLamb}) and (\ref{eq:LowBoundOnLamb}).

\section{Proof of Theorem \ref{thm:BiPartGP}}\label{sec:ProofBGP}
The proof for the bi-partite product is more complicated than the non-bipartite case. The main reason is that while there exists in the literature an exact characterization of the spectrum of $\gp(G)$~\cite{Sokarovski77}, this is not the case for $\bgp(G)$. One of the technical contributions of this paper is obtaining such a characterization, which also enables us to prove the bound on $\Lambda$ in Theorem \ref{thm:BiPartGP}.
To understand the spectrum of $\bgp(G$) we embed it in a larger graph, whose spectrum we can analyze, and then infer back the spectrum of $\bgp(G)$.

Next we define the graph in which we embed $\bgp(G)$, and state a theorem that characterizers its spectrum. Recall the definition of the \textbf{template} of an element $\ay = (a_1,\ldots,a_k) \in (X \cup Y)^k$ given in Section \ref{sec:Cospectral}: the vector $\tauu \in \{{\cal{X}},{\cal{Y}}\}^k$ that satisfies $\tau_i = \cal{X}$ iff $a_i \in X$. We are now ready to define the \emph{shuffled bi-partite graph product} in which we embed $\bgp(G)$.

\begin{definition}\label{defn:ShuffBipGraphProduct}
If $G=(X,Y,E)$ is a simple bipartite graph, define the \textbf{shuffled bi-partite graph product} $\sgp(G)$ to be the graph with vertex set $(X \cup Y)^k$, and an edge between $(a_1,\ldots,a_k)$ and $(b_1,\ldots,b_k)$ if their templates are complements of each other and there are at least $t$ pairs $(a_i,b_i)$ that share an edge in $G$.
\end{definition}

In other words, the graph $\sgp(G)$ is composed of the disjoint union of $2^{k-1}$ graphs $\bgpp{\tauu}$ for all template pairs $\tauu,\tauu^c$, amongst which is $\bgp(G)$ itself.
For example, let $G=(X,Y,E)$ be defined with $X=\{x_1,x_2\}, Y=\{y_1,y_2\}$, with only $(x_1,y_1) \in E$ and let $k=2$, $t=1$.
In this case, $\sgp(G)$ consists of the disjoint union of the two graphs $\bgpp{\tauu}$ and $\bgpp{\tauu'}$ for $\tauu = {\cal{X}}^2$ and $\tauu' = {\cal{X}}{\cal{Y}}$ (a total of $16 = |X \cup Y|^2$ vertices); while $\bgp(G) = \bgpp{\tauu}$ (a total of $8 = |X|^2+|Y|^2$ vertices).

Notice that $\sgp(G)$ differs from $\gp(G)$ (clearly it's a subgraph of $\gp(G)$, yet not an induced one). In our example, the two vertices $(x_1,y_1)$ and $(y_1,y_2)$ are connected in $\gp(G)$ but not connected in $\sgp(G)$ (as the two templates corresponding to $(x_1,y_1)$ and $(y_1,y_2)$ are not complements of each other). Therefore we have the inclusion relation $\bgp(G) \subseteq \sgp(G) \subseteq \gp(G)$.

The next theorem gives an exact characterization of the spectrum of $\sgp(G)$.
\begin{theorem}\label{thm:spectrum} Let $G=(X,Y,E)$ be a connected bi-partite $n$-vertex $d$-regular graph with eigenvalues $d=\lambda_1 \ge \lambda_2 \ge \ldots \ge \lambda_n=-d$, and corresponding eigenvectors $\uu_1,\uu_2,\ldots,\uu_n$, forming an orthonormal basis of $\R^n$. Define $\lambda^*_1=n/2-d$, $\lambda^*_n=-n/2+d$, and $\lambda^*_i = -\lambda_i$ for every $i\ne 1,n$. Let $\B$ be the set of all vectors in $\{-1,1\}^k$ having at least $t$ 1-entries. Then $\sgp(G)$ has a spectrum consisting of eigenvalues
\begin{equation}\label{eq:EigSGP}
\Lambda_{i_1i_2\cdots i_k}=\sum_{\bee \in \B} \prod_{j=1}^k \left(\frac{1+b_i}{2}\lambda_{i_j}+\frac{1-b_i}{2}\lambda^*_{i_j}\right), \qquad i_1,\ldots,i_k \in \{1,\ldots,n\}.
\end{equation}
The corresponding eigenvector is $\w_{i_1 i_2\cdots i_k}=\uu_{i_1}\otimes \uu_{i_2}\cdots \otimes \uu_{i_k}$.
\end{theorem}

Theorem \ref{thm:spectrum} is an analogue of Theorem \ref{thm:Sokarovski} from \cite{Sokarovski77}. We note that the proof of Theorem \ref{thm:spectrum}, given in Section \ref{sec:ProofSpectrum}, is similar in nature to the proof of Theorem \ref{thm:Sokarovski}. The following proposition relates the spectrum of $\sgp(G)$ and $\bgp(G)$.

\begin{proposition}\label{prop:spectrum2} Let $\sgp(G)$ be the shuffled graph product of an $n$-vertex bi-partite $d$-regular graph $G=(X,Y,E)$. Let $\Lambda_1,\ldots,\Lambda_{n^k}$ be the eigenvalues of $\sgp(G)$. Define
\begin{equation}\label{eq:SetI}
I = \{i: \Lambda_i = \Lambda_{i_1 i_2\cdots i_k} \text{ s.t. every } i_j \in \{1,n\}\}, \quad \Lambda= \max_{i \notin I}|\Lambda_i|.
\end{equation}
Then the third largest eigenvalue of $\bgp(G)$ is exactly $\Lambda$.
\end{proposition}

The proof of Proposition \ref{prop:spectrum2} is given at the end of this section and uses Theorem \ref{thm:cospectral} (co-spectrality). We are now ready to prove Theorem \ref{thm:BiPartGP}.

\bigskip

\noindent \textbf{Proof of Theorem \ref{thm:BiPartGP}:}
Proposition \ref{prop:spectrum2} gives us the exact characterization of $\Lambda$. This characterization is analogous to that of $\Lambda$ in $\gp(G)$: $\Lambda$ is the maximal among all $\Lambda_{i_1 i_2\cdots i_k}$ in which at least one $i_j$ does not correspond to any of the largest eigenvalues of $G$ (in absolute value), which are $-d$ or $d$ in the bi-partite case (i.e. $i_j \ne 1,n$). Therefore, we can bound $\Lambda$ using similar arguments to those used in the proof of Theorem \ref{thm:GP}. For the sake of completeness we give the full argument.

We start by computing the value of $\Lambda_{i_1 i_2 \cdots i_k}$ where exactly one $i_j\notin \{1,n\}$. By symmetry it doesn't matter which index it is. As for the other indices, by Theorem \ref{thm:spectrum} the choice of $i_j =1$ or $i_j = n$ only effects the sign of $\Lambda_{i_1 i_2 \cdots i_k}$, which will not matter in our case as we are interested in the absolute value. Hence we may assume w.l.o.g. that $i_1 \notin \{1,n\}$ and $i_j=1$ for $j \ge 2$.

We use similar notations to the ones in Section \ref{sec:ProofGP}: $\Psi_{s,\bee}=\prod_{j=s}^{\#\bee} \left(\frac{1+b_i}{2}d+\frac{1-b_i}{2}\left(\frac{n}{2}-d\right)\right)$, $\cC_{k,r} \subseteq \{1,-1\}^k$ is the set of all vectors with exactly $r$ one entries, and $\cC^{+}_{k,r}$ is the set of vectors with at least $r$ one entries.
Using these notations and Eq.~(\ref{eq:EigSGP}) we obtain
$$\Lambda_{i11 \cdots 1} =\sum_{\bee \in \B,b_1=1}\lambda_{i} \Psi_{2,\bee}+\sum_{\bee \in \B,b_1=-1}-\lambda_{i} \Psi_{2,\bee}=\sum_{\si \in \C^+_{k-1,t-1}}\lambda_{i} \Psi_{1,\si}-\sum_{\si \in \C^+_{k-1,t}}\lambda_{i} \Psi_{1,\si}.$$
Simplifying further we obtain the following expression for $\Lambda_{{i11 \cdots 1}}$:
\begin{equation}\label{eq:ExactFormulaBip}
\Lambda_{i11 \cdots 1}= \lambda_{i}\sum_{\si \in \cC_{k-1,t-1} }\Psi_{1,\si}.
\end{equation}
Next we upper bound $\Lambda_{i11\cdots 1}$.
\begin{align*}
d_2=\Lambda_{11\cdots 1}&=\sum_{\bee \in \B,b_1 = 1 }d\Psi_{2,\bee} + \sum_{\bee \in \B,b_1 = -1 }\left(\frac n{2}-d\right)\Psi_{2,\bee}\ge \sum_{\bee \in \B,b_1 = 1 }d\Psi_{2,\bee}\\& = \sum_{\si \in \cC^+_{k-1,t-1}} d\Psi_{1,\si} \ge \sum_{\si \in \cC_{k-1,t-1}} d\Psi_{1,\si} =\frac{d}{\lambda_i}\Lambda_{i11\cdots 1}.
\end{align*}
In other words, $\abs{\Lambda_{i11\cdots 1}} \le \frac{\abs{\lambda_i}}{d}d_2$.

The lower bound is given by Eq.~(\ref{eq:LowBoundOnLamb}) for the bi-partite case as well. Namely,
$$\abs{\Lambda_{i11\cdots 1}}\ge \frac{d_2\abs{\lambda_i}}{\alpha d}\cdot \frac{t}{k},$$
where $\alpha$ is defined in Eq.~(\ref{eq:Alpha}).

Finally, consider the case where more than one $i_j$ is greater than 1. Similar arguments to the ones above imply
that $i_1,i_2 \not \in \{1,n\}$, $\abs{\Lambda_{i_1i_211\cdots 1}}$ is smaller than $\abs{\Lambda_{i_111\cdots 1}}$.

Together with Eq.~(\ref{eq:ExactFormulaBip}) we conclude that $\Lambda = \Lambda_{i^*11\cdots 1}$ for $i^*$ s.t. $\lambda_{i^*}=\lambda(G)$, and it satisfies
$$ \frac{t}{\alpha k}\cdot \frac{\lambda}{d}\cdot d_2 \le \Lambda \le \frac{\lambda}{d} \cdot d_2.$$
This completes the proof of Theorem \ref{thm:BiPartGP}.

\subsection{Proof of Theorem \ref{thm:spectrum}}\label{sec:ProofSpectrum}
Our proof strategy, which follows the outline of the proof of Theorem 3 in \cite{Sokarovski77}, is as follows: We are going to identify the adjacency matrix $\hat A$ of the graph $\sgp(G)$, and then analyze its spectrum, and show that its eigenvectors form an orthonormal basis for $\R^{n^k}$.

We start with a few notations and definitions. Let $A$ be the adjacency matrix of the $d$-regular bi-partite graph $G=(X,Y,E)$, and let  $d=\lambda_1,\ldots,\lambda_n=-d$ be its eigenvalues with corresponding eigenvectors $\uu_1,\ldots,\uu_n$, which we may assume form an orthonormal basis for $\R^n$. Let $C$ be the following block matrix: $C_{uv}=0$ if $u,v \in X$ or if $u,v \in Y$, and  $C_{uv}=1$ otherwise (in words, $C$ is the adjacency matrix of the complete bi-partite graph). Define the bi-partite complement of $G$, denoted by $\bar{G}$, to be the bi-partite graph whose adjacency matrix is given by $\bar{A}=C-A$. For fixed $i \in [k]$ and $\bee \in \B$, define the auxiliary matrix:
$$M_{i\bee} = \frac{1+b_i}{2}A+\frac{1-b_i}{2}\bar{A}.$$
In words, $M_{i\bee}$ is either the adjacency matrix of $G$, if $b_i=1$, or of $\bar{G}$ if $b_i=-1$.

Using these notations, we claim that the adjacency matrix of $\sgp(G)$ is
\begin{equation}\label{eq:hatA}
\hat A = \sum_{\bee \in \B} M_{1\bee} \otimes M_{2\bee} \otimes \cdots \otimes  M_{k\bee}.
\end{equation}

By the definition of matrix tensor product, every entry in $\hat A$ has the form $\sum_{\bee \in \B} \prod_{i=1}^k (M_{i\bee})_{h_ig_i}$, for some $\h=(h_1,\ldots,h_k)$ and $\g = (g_1,\ldots,g_k)$ in $(X\cup Y)^k$. We address that entry in $\hat A$ by $\hat A_{\h,\g}$, namely
\begin{equation}\label{eq:hatAij}
\hat A_{\h,\g} = \sum_{\bee \in \B} \prod_{i=1}^k (M_{i\bee})_{h_ig_i}.
\end{equation}
From the definition of the tensor product, it follows that the rows and column of $\hat A$ are indexed according to the following lexicographic order on $X \cup Y$: $x_i < y_j$ for every $i,j$, $x_i < x_j$ if $i < j$ and similarly $y_i < y_j$ if $i < j$. For example, the first row of $\hat A$ is indexed by the vertex $(x_1,\ldots,x_1,x_1)$, the second row by $(x_1,\ldots,x_1,x_2)$, and so on.

Now we can show that $\hat A_{\h,\g}=1$ if $\h$ and $\g$ share an edge in $\sgp(G)$, and $\hat A_{\h,\g}=0$ otherwise. Let us start with the easier case, where the templates of $\h$ and $\g$ are not complements of each other. We need to show that $A_{\h,\g}=0$ as by the definition of $\sgp(G)$ they cannot share an edge. Since the templates are not complementary, there exists an index $i$ s.t. both $h_i$ and $g_i$ belong to, say, $X$. Therefore, no matter what $\bee \in \B$ we choose, $(M_{i\bee})_{h_ig_i}=0$, since $(h_i,g_i)$ is not an edge of $G$ nor of $\bar{G}$. Therefore the product in Eq.~(\ref{eq:hatAij}) is always zero. Next assume that $\h$ and $\g$ have complementary templates. Let $\bee$ be the vector whose entries are $b_i = 1$ if $(h_i,g_i) \in E(G)$ and $b_i=-1$ otherwise. If indeed $\h$ and $\g$ share an edge in $\sgp(G)$, then for this specific $\bee$, the value $(M_{i\bee})_{h_ig_i}$ is 1 for every $i\in [k]$, since by  definition the vector $\bee$ ``selects'' the correct matrix $A$ or $\bar A$. For all other $\bee$'s, some $(M_{i\bee})_{h_ig_i}$ is going to be zero, as for some index $i$ the ``wrong" adjacency matrix is going to be chosen. All in all, if $\h$ and $\g$ share an edge in $\sgp(G)$, then $\hat A_{\h,\g}=1$. Conversely, if $\h$ and $\g$ don't share an edge in $\sgp(G)$, then for every $\bee \in \B$, there must be some index $i$ s.t. $b_i=1$ but $(h_i,g_i)\notin E$ (otherwise, there are at least $t$ pairs $(h_i,g_i)$ that share an edge in $G$, and by the definition of $\sgp(G)$ we should have placed an edge between $\h$ and $\g$).  In this case $A_{h_ig_i}=0$ and also by definition $M_{i\bee}=A$. Thus, the product in Eq.~(\ref{eq:hatAij}) is zeroed.

Next let us show that $\w_i = \w_{i_1i_2\cdots i_k} = \uu_{i_1} \otimes \uu_{i_2} \cdots \otimes \uu_{i_k}$ is an eigenvector of $\hat A$, with eigenvalue $\Lambda_i = \Lambda_{i_1\cdots i_k}$, as defined in Eq.~(\ref{eq:EigSGP}). To that end, we first show that the eigenvalues of $\bar A$ are exactly $\lambda_1^*,\lambda_2^*,\ldots,\lambda_n^*$ as defined in Theorem \ref{thm:spectrum}, with eigenvectors $\uu_1,\ldots,\uu_n$. This follows immediately from the definition of $\bar A = C-A$, and the simple observation that $C \uu_i = 0$ for every $i \ne 1,n$ (since $\uu_i \perp \uu_1,\uu_n$, it is also perpendicular to every row of $C$ which is simply $0.5(\uu_1+\uu_n)$ or $0.5(\uu_1-\uu_n)$). To see why $\Lambda_i$ is an eigenvector of $\hat A$, use the following fact about tensor products: for two matrices $P,Q$ and two vectors $\uu,\vv$: $(P \otimes Q)(\uu \otimes \vv) = (P\uu)\otimes(Q \vv)$.

\subsection{Proof of Proposition \ref{prop:spectrum2}}
Let $G=(X,Y,E)$ be a connected bi-partite $d$-regular $n$-vertex graph, and let $\sgp(G)$ be its shuffled bi-partite graph product with vertex set $V=(X \cup Y)^k$. Set $N=2(n/2)^k$ and $M=n^k$. Recall that the different $\bgpp{\tauu}(G)$'s are cospectral (Theorem \ref{thm:cospectral}), and let us denote by $\Psi_1,\Psi_2,\ldots,\Psi_N$ their set of eigenvalues with corresponding eigenvectors $\vv_1^{\tauu},\ldots,\vv_N^{\tauu}$. Similarly, let $\Lambda_1, \Lambda_2 , \cdots , \Lambda_M$, be the eigenvalues of $\sgp(G)$.

The first observation that we make is that the spectrum of $\sgp(G)$ consists of the eigenvalues $\Psi_1,\ldots,\Psi_N$, each with multiplicity $2^k/2$. This clearly accounts for the entire spectrum of $\sgp(G)$ since $N \cdot 2^k/2 = M$. To see this, for every template pair $\tauu,\tauu^c$, embed the vector $\vv_i^{\tauu}$ in $\R^{M}$ by padding with zeros for all entries that do not correspond to $\tauu,\tauu^c$. The padded vector is an eigenvector of $\sgp(G)$ with eigenvalue $\Psi_i$. In this way we obtain $2^k/2$ linearly independent eigenvectors of $\sgp(G)$ for every eigenvalue $\Psi_i$.

The next step is to figure out the eigenvalues $\Lambda_i$ that correspond to the set $I$ defined in Eq.(~\ref{eq:SetI}). We shall use Theorem \ref{thm:spectrum} to carry out this task. Let $\lambda_1,\ldots,\lambda_n$ be the eigenvalues of $G$ with corresponding eigenvectors $\uu_1,\ldots,\uu_n$. Since $G$ is bi-partite we have $\uu_n=(\one_{n/2},-\one_{n/2})$, as well as $\uu_1=\one_n$.
Recall the definition of $\Lambda_i$ given in Eq.~(\ref{eq:EigSGP}),
\begin{equation}\label{eq:LambdaSum1}
\Lambda_i = \Lambda_{i_1\cdots i_k}= \sum_{\bee \in \B} \prod_{j=1}^k \left(\frac{1+b_j}{2}\lambda_{i_j}+\frac{1-b_j}{2}\lambda^*_{i_j}\right).
\end{equation}
By the definition of the set $I$, every $i_j$ is either 1 or $n$.
Fix $\bee \in \B$ with exactly $t$ ones.
Let $p$ denote the total number of indices $j$ s.t. $i_j=n$ in $\Lambda_i$, and let $r$ stand for the number of $j$'s s.t. $i_j=n$ and $b_j = 1$.  Substituting $\lambda_{i_j}$  and $\lambda_{i_j}^*$ according to their definition in Theorem \ref{thm:spectrum}, the product in Eq.~(\ref{eq:LambdaSum1}) equals exactly
$$d^{t-r}(-d)^r\left(-\frac{n}{2}+d\right)^{p-r}\left(\frac{n}{2}-d\right)^{k-t-(p-r)}=(-1)^p d^t \left(\frac{n}{2}-d\right)^{k-t}.$$
Summing over all vectors $\bee \in \B$ we get
\begin{equation}\label{eq:SumLambdaFinal}
\Lambda_i = (-1)^p\sum_{t'=t}^k d^{t'}\left(\frac{n}{2}-d\right)^{k-t'}\binom{k}{t'}.
\end{equation}
Recalling the value of the degree $d_2$ of $\bgp(G)$ in Eq.~(\ref{eq:deg}), we get that Eq.~(\ref{eq:SumLambdaFinal}) is exactly $(-1)^p d_2$. The set $I$ is of size $2^k$, and exactly half of the choices $i_1\cdots i_k$ have an odd number of $j$'s s.t. $i_j=n$. Therefore each of the values $d_2$ and $-d_2$ appears $2^k/2$ times in the set $\{\Lambda_i : i \in I \}$. Since $\bgpp{\tauu}(G)$ is bi-partite, we have $\Psi_1=d_2$ and $\Psi_N = -d_2$. The latter, together with the previous observation imply that the set $\{\Lambda_i : i \notin I \} = \{\Psi_2,\Psi_3,\ldots,\Psi_{N-1}\}$. In turn, $\Lambda = \max_{i \notin I} |\Lambda_i|=\max\{\Psi_2,|\Psi_{N-1}|\}$, which is the third  largest eigenvalue of $\bgp(G)$ in absolute value. This completes the proof of Proposition \ref{prop:spectrum2}.

\section{Proof of Theorem \ref{thm:cospectral}}\label{sec:ProofThmCospectral}
Before we approach the actual proof we need a few preliminary results and definitions. The following well known lemma gives a necessary and sufficient condition for two graphs to be cospectral (see for example \cite{GodsilMcKay76}, Lemma 2.1).

\begin{lemma}\label{lem:Cospec} Let $G_1$ and $G_2$ be two $n$-vertex graphs, with adjacency matrices $A_1$ and $A_2$ respectively. The two graphs $G_1$ and $G_2$ are cospectral iff $tr(A_1^\ell)=tr(A_2^\ell)$ for every $\ell \ge 1$.
\end{lemma}

Let us state the lemma in words. It is a well-known and easy-to-verify fact that the $i^{th}$ diagonal entry of the $\ell^{th}$ power of the adjacency matrix of a graph $G$ is the number of closed walks in $G$ that start and end at vertex $i$. Hence the trace of the $\ell^{th}$ power is the sum of closed walks of length $\ell$ with over counting.
In words, Lemma \ref{lem:Cospec} states that two graphs are co-spectral, if these quantities are the same in both for every $\ell$.  We shall prove a somewhat stronger claim, that in particular implies the latter. For an ordered set $\ay=(a_1,a_2,\ldots,a_\ell,a_{\ell+1})$ of vertices in $(X \cup Y)^k$, define its connectivity vector $\si(\ay)=(c_1,\ldots,c_\ell)$ according to the following rule: for every index $i$ s.t. $a_i=(u_1,\ldots,u_k)$ and $a_{i+1}=(v_1,\ldots,v_k)$ that share exactly $q$ edges $(u_s,v_s) \in E(G)$, we set $c_i = q$. In particular, $\ay$ is a walk in $\bgpp{\tauu}$ iff $c_i \ge t$ for every $i$. We say that an ordered set $(a_1,a_2,\ldots,a_\ell,a_{\ell+1})$ is \emph{circular and alternating} if $a_{\ell+1}=a_1$, and every two adjacent vertices in the set belong to complementary templates.

We shall prove the following proposition regarding the set of connectivity vectors that belong to different templates. As a simple corollary we get the cospectrality result. For simplicity, we treat all sets of vertices as ordered sets. This incurs over counting, which will not matter, as it cancels out. Also observe that two different sets of vertices may have the same connectivity vector. We treat the two vectors as different vectors. In other words, the sets $\C_{\tauu}$ defined below should be treated as {\em multi-sets}.

\begin{proposition}\label{prop:ConnectiviyVectors} Fix $\ell$, and let $\C_{\tauu}$ be the multi-set of connectivity vectors of all circular and alternating $\ell$-vertex sets in $\bgpp{\tauu}$, and similarly define $\C_{\tauu'}$. Then $\C_{\tauu'} = \C_{\tauu}$ for any pair of templates $\tauu,\tauu'$ and any $\ell \ge 0$.
\end{proposition}

Before proving the proposition, we establish the following auxiliary result concerning bi-partite $d$-regular graphs $G=(X \cup Y,E)$. We say that an ordered set $(v_1,v_2,\ldots,v_{\ell},v_1)$ obeys a vector $\uu=(u_1,\ldots,u_\ell) \in \{0,1\}^\ell$ if $(v_i,v_{i+1})$ is an edge of $G$ iff $u_i = 1$. Let $\Psi_{\uu,X}$ be the set of all alternating closed walks in $G$ that obey $\uu$ and $v_1 \in X$. Similarly define $\Psi_{\uu,Y}$.

\begin{lemma}\label{lem:NumOfTemplatePathsInG} Let $G=(X \cup Y,E)$ be a $d$-regular $n$-vertex bi-partite graph. For every $\ell \ge 0$ and every vector
$\uu=(u_1,\ldots,u_\ell) \in \{0,1\}^\ell$, we have $|\Psi_{\uu,Y}|=|\Psi_{\uu,X}|$.
\end{lemma}
\begin{Proof}
If $\ell$ is odd, then clearly $|\Psi_{\uu,Y}|=|\Psi_{\uu,X}|=0$, since $G$ is bi-partite. Now consider an even $\ell$. Let $C$ be the adjacency matrix of the complete bi-partite $n$-vertex graph, and let $A$ be the adjacency matrix of $G$. Recall our definition of the bi-partite complement of $G$, denoted by $\bar{G}$, which is the bi-partite graph whose adjacency matrix is given by $\bar{A}=C-A$. Let $I_1$ be the $n \times n$ matrix, whose first $n/2$ diagonal entries are 1, and all other entries are 0. Similarly define $I_2$ w.r.t. to the last $n/2$ diagonal entries. Using these definitions we claim that
\begin{equation}\label{eq:PsiXPsiY}
|\Psi_{\uu,X}|=tr\left( I_1 \cdot \prod_{i=1}^{\ell}\left(u_i A + (1-u_i) \bar{A}\right) \right),\qquad |\Psi_{\uu,Y}|=tr\left( I_2 \cdot \prod_{i=1}^{\ell}\left(u_i A + (1-u_i) \bar{A}\right) \right).
\end{equation}
To see this, note that the $j^{th}$ diagonal entry in the matrix product $\prod_{i=1}^{\ell}\left(u_i A + (1-u_i) \bar{A}\right)$ is the total number of closed walks that start at $x_j$ (if $j \le n/2$, or $y_{j-n/2}$ if $j > n/2$) and obey $\uu$. If $\uu = 1^\ell$ for example, then we get the standard interpretation of taking powers of the adjacency matrix. Multiplying by $I_1$ (or $I_2$) and taking the trace simply extracts the relevant paths: starting at $X$ or $Y$ (the trace of a matrix product obeys the rule $tr(A^T B) = \sum_{i,j}A_{ij}B_{ij}$).

The product $\prod_{i=1}^{\ell}\left(u_i A + (1-u_i) \bar{A}\right)$ defines a polynomial $P$ in $A$ (replace $\bar{A}$ by $C-A$). When simplifying the product, we get two types of terms: we get $A^\ell$, or terms of the form $(-1)^{s_2+s_4+...} A^{s_1}C^{s_2}A^{s_3}\cdots $ for some natural numbers $s_i$. To proceed we observe that $(a)$ $A$ and $C$ commute $(b)$ the product $AC$ equals $d(J-C)$, where $J$ is the all-one matrix and $C$ is the adjacency matrix of the complete bi-partite graph, and $(c)$ the product $(J-C)A=dC$ and $(J-C)C=\frac{n}{2}C$. Using these three observations one can readily verify that the second type of terms simplify to $b \cdot (J-C)$, for $b=d^{s_1+s_3+...}\left(\frac{n}{2}\right)^{-1+s_2+s_4+...}$. Let us write the polynomial $P$ as $P=A^\ell + T$, where $T$ is the sum of all type-two terms. Using the linearity of the trace operator, Eq.~(\ref{eq:PsiXPsiY}) can be rewritten as $tr\left( I_1 \cdot \prod_{i=1}^{\ell}\left(u_i A + (1-u_i) \bar{A}\right) \right)=tr(I_1 A^{\ell}) + tr(I_1 T).$
By our characterization of $T$, it holds that $tr(I_1 T)=  tr(I_2 T)$. As for $tr(I_1 A^{\ell})$, this is the sum over all $i$ of the number of closed walks of length $\ell$ that start at $x_i$. The same is true for $tr(I_2 A^{\ell})$ just with $y_i$. However a cyclic shift is a natural bijection between the two sets, namely map $x_{i_1}-y_{i_1}-x_{i_2}- ...$ to $y_{i_1}-x_{i_2}-...$.
To conclude, we got
\begin{align*}
|\Psi_{\uu,X}|&=tr\left( I_1 \cdot \prod_{i=1}^{\ell}\left(u_i A + (1-u_i) \bar{A}\right) \right)=\\& =tr(I_1 T)+tr(I_1 A^{\ell})=
tr(I_2 T)+tr(I_2 A^{\ell})=tr\left( I_2 \cdot \prod_{i=1}^{\ell}\left(u_i A + (1-u_i) \bar{A}\right) \right)=|\Psi_{\uu,Y}|.
\end{align*}
\end{Proof}

We are now ready to prove Proposition \ref{prop:ConnectiviyVectors}

\medskip

\begin{Proof}(Proposition \ref{prop:ConnectiviyVectors})
We fix the parameter $\ell$ and prove via induction on $k$. The base case, $k=1$, is true in a vacuous way since for $k=1$ there is only one template pair.
Assume the claim is true up to $k-1$ and let us prove it for $k$.
Fix a template $\tauu$ of length $k$, and let $\hat \tauu$ be its $(k-1)$-prefix. The graph $BGP_{k,t,\tauu}$ is obtained from $BGP_{k-1,t,\hat \tauu}$ by extending each $(k-1)$-tuple with a vertex from $X \cup Y$ (according to $\tauu$) and updating the edge set accordingly (existing edges remain, but new edges may appear).

Let $\si \in \C^{(k-1)}_{\hat \tauu}$ be a connectivity vector corresponding to the (circular and alternating) ordered set $\ay=(a_1,a_2\dots,a_\ell,a_{\ell+1})$ with each $a_i$ in $BGP_{k-1,t,\hat \tauu}$.
Let $\uu \in \{0,1\}^\ell$.
An update of $\ay$ by $\uu$ is an {\em extended} version $\ay^{(k)}$ of $\ay$ in which one appends to each $a_i \in (X \cup Y)^{k-1}$ a vertex
$v_i$ in $X \cup Y$ (according to the template $\tauu$) such that $v=(v_1,v_2,\dots,v_\ell,v_{\ell+1})$ in an alternating closed walk that obeys $\uu$, i.e., $v$ is either in $\Psi_{\uu,X}$ or $\Psi_{\uu,Y}$.
For any such extension of $\ay^{(k)}=(a^{(k)}_1,\dots,a^{(k)}_\ell,a^{(k)}_{\ell+1})$ (which is circular and alternating and consists of $k$-tuples $a^{(k)}_i \in (X \cup Y)^k$ of $BGP_{k,t,\tauu}$), the corresponding connectivity vector is exactly $\si+\uu$.
The set $\C_{\tauu}$ consists of the multi-set of connectivity vectors obtained by this process starting from any $\ay$ and any $\uu$ as above.

To prove the induction step, it suffices to show that for any  connectivity vector $\si \in \C^{(k-1)}_{\hat \tauu}$ corresponding to a circular and alternating $\ay=(a_1,a_2\dots,a_\ell,a_{\ell+1})$ in $BGP_{k-1,t,\hat \tauu}$ and any vector $\uu \in \{0,1\}^\ell$, the number of extensions $\ay^{(k)}$ of $\ay$ by $\uu$ does not depend on $\ay$, $\tauu$ or $\hat \tauu$.

The latter follows from the fact that the number of alternating closed walk $v=(v_1,v_2,\dots,v_\ell,v_{\ell+1})$ that obey $\uu$ is determined completely by $G=(X,Y,E)$ and is independent of $\ay$ and $\hat \tauu$.
However, the value at hand may potentially depend on $\tauu$ only through the question whether $v_1 \in X$ or $v_1 \in Y$. For $v_1 \in X$ the number of alternating closed walks is the size of $\Psi_{\uu,X}$ and for $v_1 \in Y$, the size of $\Psi_{\uu,Y}$.
By Lemma \ref{lem:NumOfTemplatePathsInG} these two are equal and we may conclude our proof.
%
%
%
%
\end{Proof}

\section{Proof of Proposition \ref{prop:deg}}\label{sec:degProof}
Fix a vertex $\x =  (x_1,\ldots,x_k)$, and let us compute the number of neighbors $(y_1,\ldots,y_k)$ that $\x$ has in $\gp(G)$. Fix some $t' \geq t$, there are  $\binom k{t'}$ ways to choose the indices $j$ for which $(x_j,y_j)\in E(G)$. For every such $j$, there are $d$ possible $y_j$'s. For the other $x_j$'s, there are $n-d$ non-neighbors to choose from (this includes $x_j$ itself, as the graph $G$ is simple). Now sum over all possible $t' \geq  t$ to obtain
$$d_1=\sum_{t'=t}^k \binom {k}{t'}d^{t'}(n-d)^{k-t'}.$$

Since $\x$ was arbitrary, this equality holds for every $\x$, hence $\gp(G)$ is $d_1$-regular. Similar arguments give the bi-partite case, and the shuffled product, which is composed of $2^{k-1}$ disjoint bi-partite graphs, similar to $\bgp(G)$.

\end{document}